\documentclass[aps,prl,preprint,superscriptaddress,floatfix]{revtex4-1}%

\usepackage{siunitx}

\usepackage{times}

\usepackage[utf8x]{inputenc}
\usepackage[english]{babel}
\usepackage[T1]{fontenc}
\usepackage{lmodern}
\usepackage{amssymb}
\usepackage{amsmath}
\usepackage{amsthm}
\usepackage[pdftex]{graphicx}
\usepackage{epstopdf}
\usepackage{braket}
\usepackage[bottom]{footmisc}

\usepackage{dcolumn}
\usepackage{bm}
\usepackage{color}
\usepackage{relsize,dsfont,mathrsfs,empheq,verbatim,upgreek}
\usepackage{etoolbox}
\usepackage[caption = false]{subfig}
\usepackage{multirow}
\usepackage{physics}

\usepackage{hyperref}

\newcommand{\detuning}{\Delta}

\newcommand{\T}{T}

\newcommand{\omegaTimeAverage}{\langle \delta \tilde{\omega}(t) \rangle_{\T}}
\newcommand{\omegaTimeAverageNoShift}{\langle \tilde{\omega}(t) \rangle_T}
\newcommand{\omegaTimeDependent}{\delta\tilde{\omega}(t)}
\newcommand{\nth}{\bar{n}}

\newcommand{\omegam}{\omega_m}

\newcommand{\analysisphase}{\varphi}

\newcommand{\Beff}{\mathbf{B}} 
\newcommand{\accumulatedPhase}{\tilde{\varphi}_T}
\renewcommand{\S}{S}
\newcommand{\E}{E}
\newcommand{\g}{g}

\newcommand{\detuningBeff}{\detuning(\Beff)}
\newcommand{\sigmay}{\langle \sigma_y \rangle}
\newcommand{\LDp}{\eta}
\newcommand{\Tdetuning}{\T(\detuning)}
\newcommand{\sgn}{\mathrm{sgn}}

\newcommand{\omegaasterisk}{\omega^\ast}

\newcommand{\Figa}{a}
\newcommand{\Figb}{b}
\newcommand{\Figc}{c}

\addto\captionsenglish{}

\addto\captionsenglish{}

\begin{document}

\title{Observing Time-Dependent Energy Level Renormalisation\\ in an Ultrastrongly Coupled Open System}


\author{Alessandra Colla}

\affiliation{Institute of Physics, University of Freiburg, 
Hermann-Herder-Stra{\ss}e 3, D-79104 Freiburg, Germany}

\affiliation{Dipartimento di Fisica ``Aldo Pontremoli'', Universit\`a degli Studi di Milano, Via Celoria 16, I-20133 Milan, Italy}

\author{Florian Hasse}

\affiliation{Institute of Physics, University of Freiburg, 
Hermann-Herder-Stra{\ss}e 3, D-79104 Freiburg, Germany}

\author{Deviprasath Palani}

\affiliation{Institute of Physics, University of Freiburg, 
Hermann-Herder-Stra{\ss}e 3, D-79104 Freiburg, Germany}

\author{Tobias Schaetz}

\affiliation{Institute of Physics, University of Freiburg, 
Hermann-Herder-Stra{\ss}e 3, D-79104 Freiburg, Germany}

\affiliation{EUCOR Centre for Quantum Science and Quantum Computing,
University of Freiburg, Hermann-Herder-Stra{\ss}e 3, D-79104 Freiburg, Germany}

\author{Heinz-Peter Breuer}

\affiliation{Institute of Physics, University of Freiburg, 
Hermann-Herder-Stra{\ss}e 3, D-79104 Freiburg, Germany}

\affiliation{EUCOR Centre for Quantum Science and Quantum Computing,
University of Freiburg, Hermann-Herder-Stra{\ss}e 3, D-79104 Freiburg, Germany}

\author{Ulrich Warring}

\affiliation{Institute of Physics, University of Freiburg, 
Hermann-Herder-Stra{\ss}e 3, D-79104 Freiburg, Germany}

\begin{abstract}
Understanding how strong coupling and memory effects influence the energy levels of open quantum systems is a complex and challenging problem.
Here, we show these effects by probing the transition frequency of an open two-level system within the Jaynes-Cummings model, experimentally realised using Ramsey interferometry in a single trapped $^{25}$Mg$^{+}$ ion. 
Measurements of the system, coupled to a single-mode environment, reveal a time-dependent shift in the system's energy levels of up to 15\% of the bare system frequency. 
This shift, accurately predicted using an open system ansatz of minimal dissipation, results purely from ultra-strong system-mode interactions and the buildup of correlations. 
Time-averaged measurements converge to the dispersive Lamb shift predictions and match dressed-state energies, indicating that this observed shift represents a generalised Lamb shift applicable across all coupling and detuning regimes.
Our findings provide direct evidence of dynamic energy level renormalisation in strongly coupled open quantum systems, although the total system-environment Hamiltonian is static; this underscores the significance of memory effects in shaping the reduced system's energy landscape. 
These results offer more profound insights into Hamiltonian renormalisation, essential for strong-coupling quantum thermodynamics and advancements in all quantum platforms.
\end{abstract}

\maketitle

Stable control and manipulation are critical for quantum technology, requiring a deep understanding of the underlying interactions of physical platforms with their surroundings.
While open system theory has dramatically advanced as a tool for predicting the evolution of a quantum system coupled to a general environment~\citep{Breuer2007}, more profound questions, e.g., how the energy of the system changes due to coupling to its environment, especially when the interaction is strong, and memory effects are present, are still lacking satisfactory answers.

\begin{figure}
    \centering
    \includegraphics[]{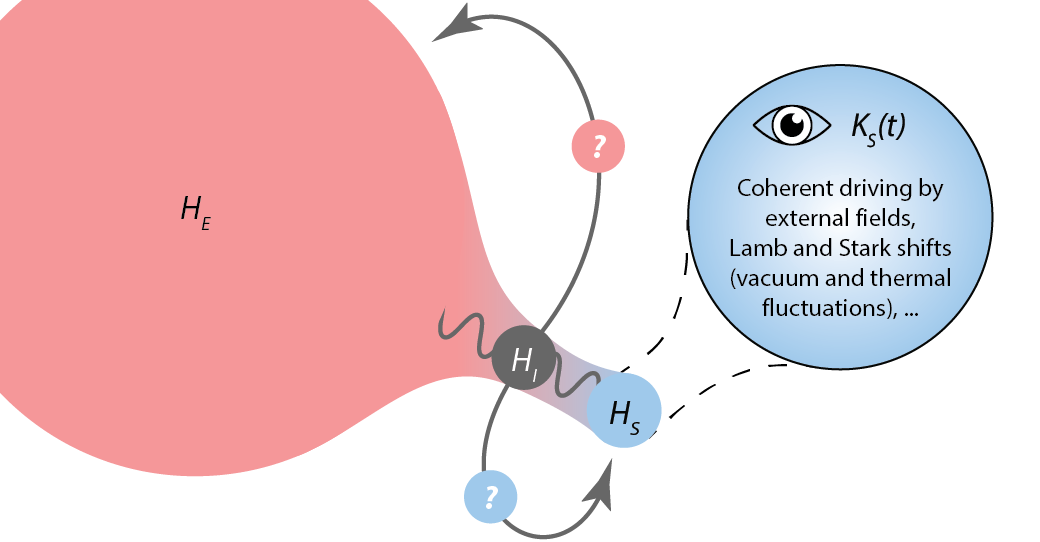}
    \caption{\textbf{Renormalisation of an open system Hamiltonian due to averaging out the environmental degrees of freedom.}
    In an open quantum system (Hamiltonian $H_S$), strong or structured interactions (Hamiltonian $H_I$) with some environment (Hamiltonian $H_E$) can lead to memory effects and enforce non-negligible energy contributions. The interaction affects, in principle, both the system's reduced dynamics and energy. We investigate an open-system approach to describe the renormalisation of the system energy levels when strongly interacting with a non-Markovian environment. The renormalisation is generally time-dependent and assumed to be uniquely encoded in the emergent Hamiltonian $K_S(t)$ via the principle of minimal dissipation. Examples of known renormalisation in light-matter interaction derived from this approach are the Lamb and AC Stark shifts (for time-independent shifts) and semi-classical coherent driving (for time-dependent shifts).
    We validate this approach outside the known Markovian regimes by performing measurements on the open system only, providing strong evidence of the uniqueness of the effective system Hamiltonian in a non-Markovian scenario.
    }
    \label{fig1}
\end{figure}

Short-range interactions and scaling arguments justify neglecting the energy contribution associated with the system-environment interaction for a macroscopic classical system interacting with its surroundings. However, for a quantum system, the interaction energy can be of comparable magnitude to that of the system. Thus, it can be expected to influence its energy and dynamics, mainly when the environment is finite, structured or for strong couplings. Since, formally, the interaction energy is shared between the system and the environment, it is unclear how much of this energy needs to be attributed to the system, cp. Fig.\,\hyperref[fig1]{1}. Consequently, predictions of open systems' energy levels are not uniquely defined but play a crucial role in the design and control of quantum technology platforms~\cite{Freer2017,Takeda2018,Hendrikx2020,Philips2022}.
Furthermore, determining the system's energy in strong coupling scenarios can provide insights on quantities like work and heat in non-equilibrium quantum thermodynamics~\cite{Kosloff2013,Alicki2018_Springer_Version}, a field where existing theoretical approaches offer conflicting results~\cite{Weimer2008,Esposito2010,Teifel2011,Alipour2016,Seifert2016,Strasberg2017,Rivas2020,Alipour2021_PRA,Landi2021,Elouard2023,Seegebrecht2023_Springer}.

A dynamically inspired, out-of-equilibrium approach to this question~\cite{Colla2022} predicts an effective energy level renormalisation due to environmental interactions. The effect generally depends on the parameters determining the total Hamiltonian, particularly the coupling strength and details of the initial environmental states, such as its temperature. Known examples of the energy level renormalisation described by~\cite{Colla2022} for weakly coupled Markovian environments are given by the Lamb and AC Stark shifts induced in atomic energy levels by the vacuum and thermal fluctuations of the electromagnetic field~\cite{Lamb1947,Autler1955,Breuer2007}.
These effects are captured by the Jaynes-Cummings (JC) model in the dispersive limit, where a two-level system is weakly coupled to a single bosonic mode with large detuning. They have been experimentally measured in a wide coupling range, e.g., from Rydberg atoms~\cite{Brune1994} to superconducting qubits~\cite{Fragner2008} coupled to microwave cavities.

The JC model is indeed prototypical as it describes not only physical situations in traditional cavity quantum electrodynamics, where a two-level atom is interacting with an optical or microwave cavity but also in several other modern experimental platforms -- where the traditional constituents are replaced by analogue versions -- such as quantum dots, superconducting qubits and trapped ions~\cite{USC_RMP_2019}.
Furthermore, by identifying the two-level system as the system and the bosonic mode as the environment, the JC model provides a fundamental playground to investigate the less understood strong coupling and non-Markovian regimes, e.g., in trapped-ion experiments~\cite{Gessner2014, Wittemer2018}, which emerge at higher values of coupling strength and close to resonance.
In such scenarios, the predicted effective energy levels of the system might change with the coupling duration, describing an effective driving of the open system energy levels even if the total system-environment-interaction Hamiltonian is time-independent~\cite{Colla2022}. 
In quantum thermodynamics, this is interpreted as effective work, emerging from tracing the degrees of freedom of the environment~\cite{Picatoste2023}.

In this article, we investigate the influence of strong coupling and memory effects on the energy levels of open quantum systems by probing the transition frequency of an open two-level system within the JC model, experimentally realised by a single trapped  $^{25}$Mg$^{+}$ ion.
By conducting measurements near resonance between a two-level system and a single motional mode (pure quantum environment), we observe a time-dependent shift in the system's energy levels of up to 15\% of the bare spin frequency, despite the absence of external driving protocols.
We interpret this effect as an emergent driving of the system resulting from its coupling with the environment.
The observed shift is accurately predicted using the principle of minimal dissipation~\cite{Colla2022}; we additionally recover the Lamb shift and dressed-state energies when performing time-averaged measurements and thus identify this shift as a generalised Lamb shift, valid for all coupling regimes and even near resonance.
These measurements contribute to a deeper fundamental understanding of Hamiltonian renormalisation in open quantum systems, setting the stage for the exploitation of time-dependent energy level shifts. This can have significant consequences for strong-coupling quantum thermodynamics, where energy shifts are often neglected, and for virtually all quantum technology platforms.

According to the approach of Ref.~\cite{Colla2022}, the system's energy is described by an effective, potentially time-dependent Hamiltonian operator $K_S(t)$. This emergent Hamiltonian is a consequence of the joint dynamics of the system $\S$ and the environment $\E$ and can be observed by probing solely $\S$, see Fig.\,\hyperref[fig1]{1}. We find $K_S(t)$ within the time-local master equation, which describes the exact evolution of $\S$ and is formally obtained with no approximations by tracing out the degrees of freedom of $\E$.
In very general terms, the master equation is written as a sum of two terms, namely a commutator with the Hamiltonian $K_S(t)$ and a dissipator term~\cite{Gorini1976, Hall2014, Breuer2012}
\begin{eqnarray} \label{tcl-meq}
 \frac{d}{dt}\rho_S(t) &=&-i \left[K_S(t),\rho_S(t)\right] \\ \nonumber &+&\sum_{k}\gamma_{k}(t)\Big[L_{k}(t)
  \rho_S L_{k}^{\dag}(t) - \frac{1}{2}\big\{L_{k}^{\dag}(t)L_{k}(t),\rho_S\big\}\Big],
\end{eqnarray}
where the time-dependent rates $\gamma_k \in \mathbb{R}$ can be negative. A master equation in this form (also sometimes referred to as \textit{generalised Lindblad form}) axiomatically exists for all open system evolutions, except for at most a set of discrete points on the time axis~\cite{Breuer2012, Stelmachovic2001}, even if $\S$ and $\E$ are initially correlated~\cite{Colla2022b}. It is also already known exactly for a set of integrable models -- e.g., for systems linearly coupled to bosonic environments~\cite{Hu1992,Jin2010,Lei2012,Ferialdi2016}, various dephasing models~\cite{Tu2008} and the JC model~\cite{Smirne2010}.  
When no exact treatment is available, the master equation can still be evaluated numerically or approximated through a time-convolutionless perturbation expansion~\cite{Shibata1977,Chaturvedi1979,Gasbarri2018} that can show leading strong coupling and memory effects, and that can still be put in the form of equation \eqref{tcl-meq}. 

The splitting between a commutator and a dissipator in the master equation, and thus the exact expression for the Hamiltonian $K_S$, is, however, highly non-unique~\cite{Breuer2007,Chruscinski2022_PhysicsReports}. Nonetheless, given any open system evolution, the splitting can be uniquely assigned by imposing a recent minimal dissipation principle~\cite{Sorce2022} that requires minimisation of the dissipator's action as a superoperator by averaging over input and output states to recognise the part of the evolution which is coherent and retrievable by $\S$. This is taken in~\cite{Colla2022} to be an additional physical principle that uniquely specifies the emergent Hamiltonian $K_S(t)$ and, consequently, the system's modified energy levels, reproducing, e.g., semi-classical coherent driving~\cite{Colla2024thesis} and the known Lamb and AC Stark shifts~\cite{Breuer2007} in the appropriate cases. Coherent driving is one example of a time-dependent emergent Hamiltonian, which is generally true for non-Markovian environments, even if the total Hamiltonian is time-independent~\cite{Picatoste2023}.

\begin{figure}
  \includegraphics[]{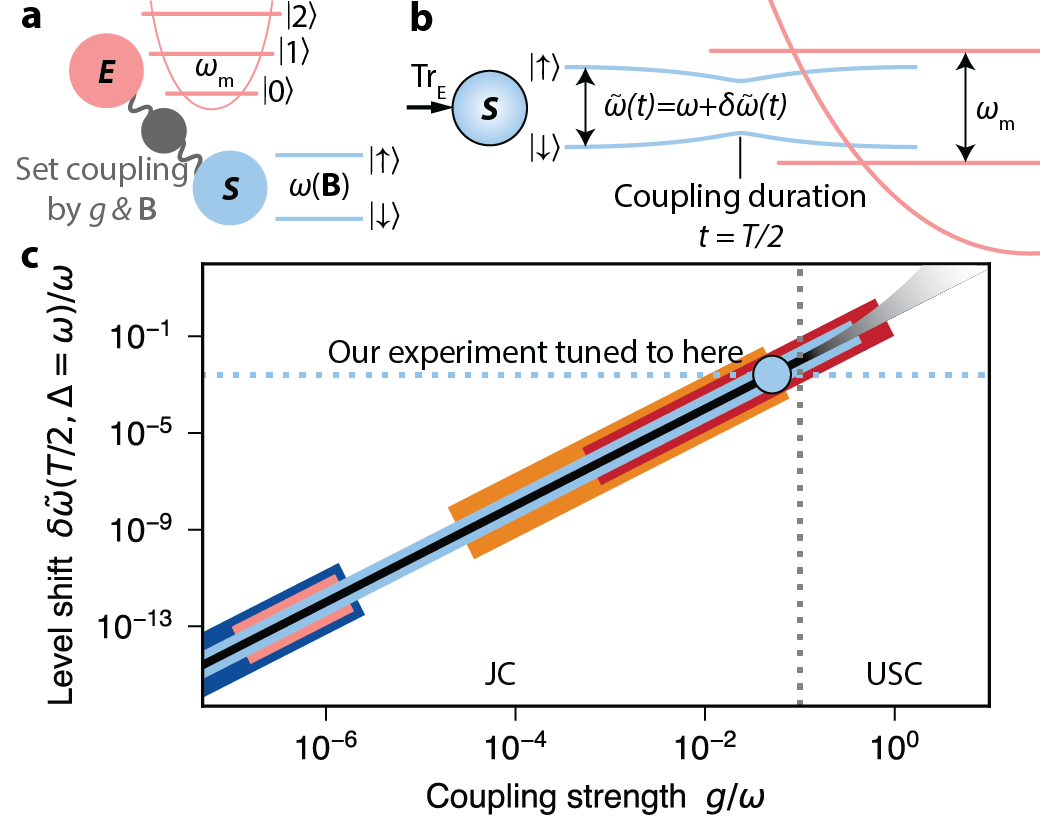}
    \caption{\textbf{Perspective on paradigmatic open-system dynamics, validating the time-dependence of significant energy renormalisation in the Jaynes-Cummings (JC) model near the ultrastrong coupling (USC) threshold.}
    \textbf{\Figa}, Experimentally, we utilise a single bosonic mode (cavity analogue) with eigenfrequency $\omegam$ of a trapped magnesium cation as environment $E$ and, as system $S$, the dressed electronic states (atom analogue) ($\ket{\uparrow}$, $\ket{\downarrow}$) with bare spin frequency $\omega(\Beff)$ tuned by a static effective magnetic field $\Beff$ (Zeeman-shift analogue).
    We can continuously tune the $S$—$E$ coupling strength $g$ near the USC threshold.
    \textbf{\Figb}, Following~\cite{Leibfried2003}, we simulate the cavity quantum electrodynamics analogue of the JC model with the dressed magnesium ion in the red-sideband regime.
    Ref.~\cite{Colla2022} predicts a significant energy shift as a function of coupling duration $t$ with a maximum level shift $\delta\tilde{\omega}(T/2) = -\frac{2g^2}{\detuning}$ for the mode initially in the motional ground state~\cite{Supplement}.
    This effect can be resonantly enhanced for small detuning $\detuningBeff=\omegam-\omega(\Beff)$.
    \textbf{\Figc}, The shift is estimated in the JC model to be $\propto g^2$ and is shown as a function of $g$ (black line) for several orders of magnitude.
    We emphasise the significance for different experimental platforms~\cite{USC_RMP_2019}: Atoms in optical (dark blue) and microwave (light red) cavities, quantum dots (orange), superconducting qubits (red) and trapped atomic ions (light blue).
    The target regime of our work (light blue disk) is near the USC.
    Beyond the USC, the JC approximation loses its validity toward the deep strong coupling.}\label{fig2}
\end{figure}

As depicted in Fig.\,\hyperref[fig2]{2a}, in the following, we consider $S$ given by a two-level system, which is coupled to a single bosonic degree of freedom acting as $E$, described by the JC Hamiltonian
\begin{equation}\label{JC_ham}
H = \frac{\hbar \omega} {2} \sigma_z + \hbar \omega_m a^{\dag}a + \hbar g (\sigma_+ a + \sigma_- a^{\dag}) \;,
\end{equation}
where $\sigma_{z}$ is the Pauli $z$-operator, $\sigma_{+}$ and $\sigma_{-}$ and $a^\dagger$ and $a$ are the raising and lowering operators of the spin and the mode, respectively. The bare spin and mode frequencies are given by $\omega$ and $\omega_m$, respectively, while $g$ denotes the coupling parameter, and $\hbar$ is the reduced Planck constant.

For an initial thermal state of the mode, described by mean excitation number $\nth=(e^{\beta \hbar \omegam}-1)^{-1}$ with $\beta$ its initial inverse temperature, the exact time-convolutionless master equation for $S$, in generalised Lindblad form equation \eqref{tcl-meq} and in minimal dissipation, reads~\cite{Smirne2010}:
\begin{eqnarray}\label{JCme}\nonumber
\dot{\rho}_S = &-& i \left[\frac{\tilde{\omega}(t)}{2}\sigma_z, \rho_S\right] + \gamma_z(t) \left[ \sigma_z \rho_S \sigma_z - \rho_S\right] \\ \nonumber
 &+& \gamma_+(t) \left[ \sigma_+ \rho_S \sigma_- - \frac{1}{2}\{ \sigma_- \sigma_+, \rho_S \} \right]  \\ 
&+& \gamma_-(t) \left[ \sigma_- \rho_S \sigma_+ - \frac{1}{2}\{ \sigma_+ \sigma_-, \rho_S \} \right]
\end{eqnarray}
with time-dependent coefficients $\tilde{\omega}, \gamma_{+,-,z}$ depending on $\bar{n}$ and on the Hamiltonian parameters, $g$ and the spin-mode detuning $\detuning = \omega_m-\omega$. 
The master equation \eqref{JCme} reveals an emergent Hamiltonian of the form \mbox{$K_S(t)=\tilde{\omega}(t)\sigma_z/2$}, where the transition frequency between ground and excited state is renormalised via a time-dependent shift $\omegaTimeDependent$, i.e., $\tilde{\omega}(t) = \omega + \omegaTimeDependent $.
The dependence on the coupling duration is due to the mode environment's finite size and the related presence of memory effects. 
It can be interpreted as an energy exchange between the system and the environment, which appears as a driving of the system, as schematically illustrated in Fig.\,\hyperref[fig2]{2b}.
The average magnitude of the renormalisation is proportional to the square of $g$~\cite{Supplement} and can, for most platforms, be described by the JC model.
Usually, the energy level shift remains several orders of magnitude smaller than the bare spin energy splitting; see Fig.\,\hyperref[fig2]{2c}.
But trapped ions and superconducting qubits can tap into the ultrastrong coupling (USC) regime~\cite{USC_RMP_2019}, where the renormalisation contribution reaches about 10-20$\%$ of $\omega$.

The transition frequency of $\S$, and thus its (possibly time-dependent) energy level splitting, can be detected via its Larmor frequency. In our case, it holds analytically that the Larmor frequency is identical to the renormalised energy splitting in the Hamiltonian $K_S(t)$~\cite{Smirne2010}.
While this is true for any initial temperature of the mode, the special case of the mode initially in the vacuum gives an analytic, explicit expression for $\omegaTimeDependent$, which is periodic in time with the $S$-$E$ coupling duration \mbox{$\Tdetuning = 2\pi/\sqrt{\detuning^2 + 4\g^2}$}, and reads
\begin{equation}\label{Eq_shift}
   \omegaTimeDependent = - \frac{2g^2}{\Delta} \frac{1}{1+ \left(1+ \frac{4\g^2}{\Delta^2}\right) \cot^2\left(\frac{\pi}{T(\Delta)} t \right)} \; .
\end{equation}
In this zero-temperature case, the average shift is given by
\begin{equation}\label{Eq_average_shift}
    \langle \delta \tilde{\omega}(t) \rangle_{\Tdetuning} = \frac{-2 g^2 \sgn(\detuning)}{\lvert \detuning \rvert + \frac{2\pi}{\Tdetuning}} \;,
\end{equation}
and it is identical to the dressed-state energy of the total Hamiltonian equation \eqref{JC_ham} associated with the excited spin~\cite{Supplement}.

We implement the model experimentally with a single trapped magnesium cation, choosing $E$ represented by a single motional mode with $\omegam \simeq 2\pi 1.3$\,MHz, while the energy of $S$ is tuned, by an effective static magnetic field $\Beff$ (analogue to a Zeeman shift), near  $\omega(\Beff)\approx \omegam$, cp. Fig.\, \hyperref[fig2]{2a}.
We ensure near identical dynamics as JC model and perform the experiments in the so-called red-sideband regime~\cite{Leibfried2003, Supplement}. 
Further, we perform numerical simulations of the trapped-ion (TI) Hamiltonian neglecting technical limitations \cite{Supplement, clos_time-resolved_2016, Wittemer2018, Hasse_2024_Strobo}.
In-depth information about the experimental setup, data recording, and analysis are given in Methods. In contrast, numerical simulations and comparisons of the JC approximation to the red-sideband regime are given in the Supplement~\cite{Supplement}.

\begin{figure}
  \includegraphics[]{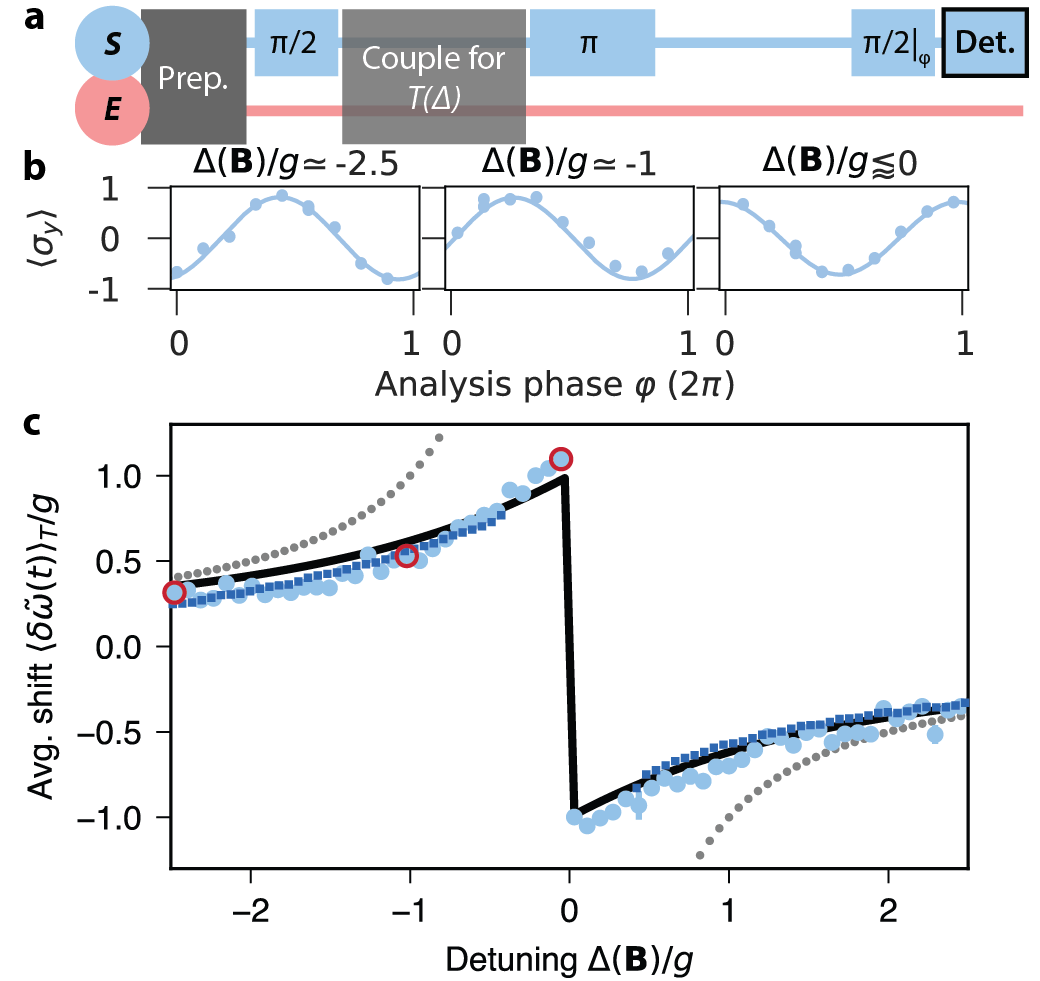}
    \caption{\textbf{Experimental probing of time-averaged energy renormalisation for narrow detuning range} 
    \textbf{\Figa}, Experimental sequence: The motional mode ($E$) with frequency \mbox{$\omega_m = 2\pi\,1.303(1)$ MHz} is cooled close to the motional ground state with a residual thermal contribution of \mbox{$\nth = 0.08(3)$}.
    A first $\pi/2$-pulse initialises the spin ($S$) in a superposition state with equal weights.
    The $S$-$E$ coupling pulse with a coupling strength of $g\simeq2\pi\,0.065$ MHz is applied for a duration of $\T(\detuning)$, followed by a spin flip ($\pi$-pulse) and a free evolution matching $\T(\detuning)$.
    A second $\pi/2$-pulse with variable analysis phase $\analysisphase$ is applied, succeed by fluorescence detection.
    \textbf{\Figb}, Raw data (light blue disks) with sinusoidal fit (light blue solid line) for three distinct detunings $\detuningBeff/g$.
    Every data point consists of 100 experimental repetitions of the same parameter settings.
    The phase change is proportional to the time-average shift.
    A detailed description of the analysis used is given in the Methods section.
    Error bars are smaller than the data points.
    \textbf{\Figc}, Following the analysis procedure, we show experimental results (light blue disks); red circles mark results from the selected raw data from \Figb\.
    For $\detuning>0$ the data points match the JC prediction, equation (\ref{Eq_average_shift}), (black solid line) and TI simulation (dark blue dotted line), as well as, for $\detuning \gg g$ agreement with the Lamb-shift (grey dotted line).
    All error bars represent the standard error of the mean.}
    \label{fig3}
\end{figure}

\begin{figure}[]
\includegraphics[]{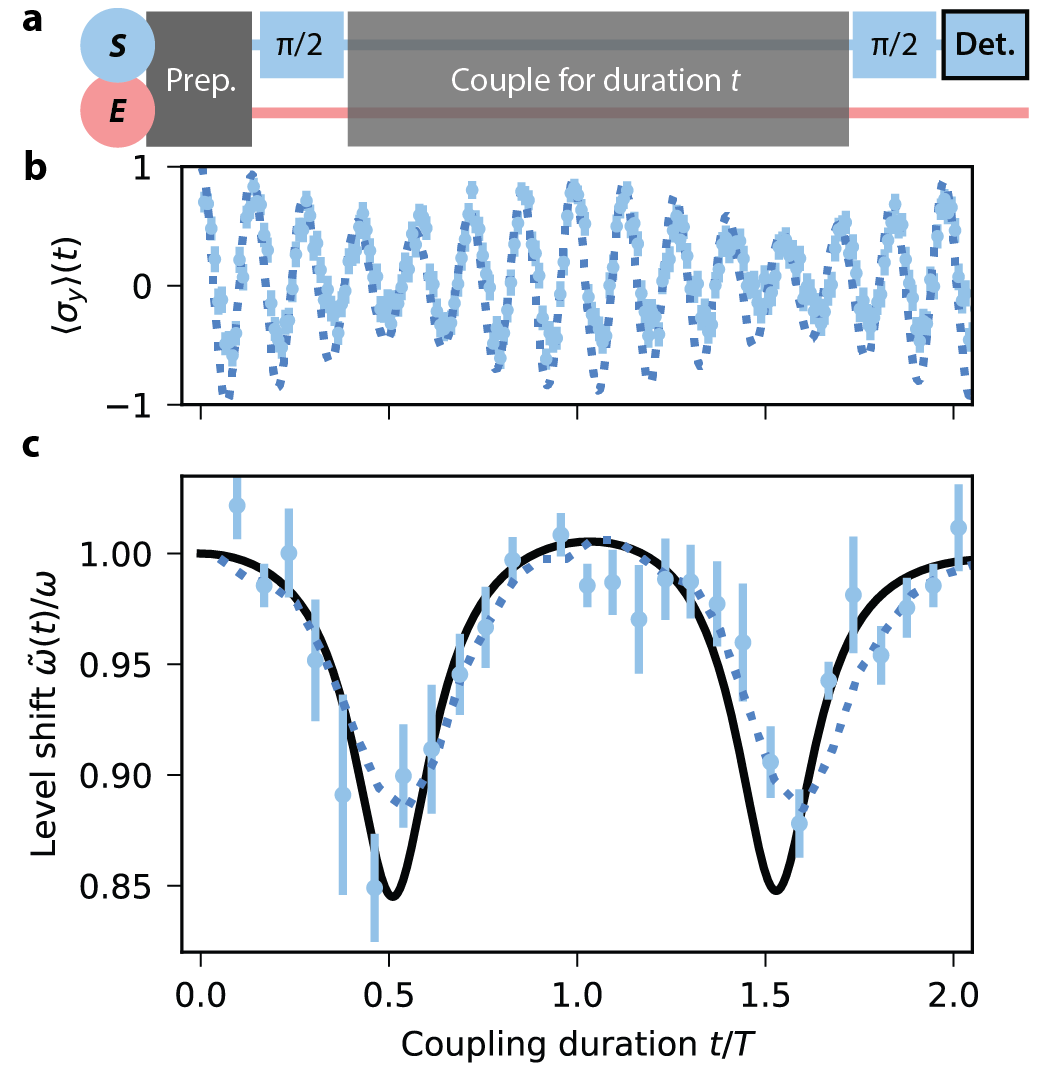}%
    \caption{
    \textbf{Observing time-resolved Larmor frequency modulation as a proof of time-dependent energy renormalisation near the USC regime.} 
    \textbf{\Figa}, In our experiment, we initialise the spin ($S$) in equal superposition of $\ket{\uparrow}$ and $\ket{\downarrow}$, and the mode ($E$) with a frequency of $\omega_m = 2\pi\,1.304(1)$ MHz near the vacuum state with a residual thermal contribution of $\nth = 0.08(2)$. 
    We realise $\omega = 2\pi\,1.24(3)$\,MHz (free Larmor frequency), and coherently couple the spin to the mode for variable duration with the coupling strength $g \simeq 2\pi\,0.078(2)$~MHz, resulting in a fixed detuning $\detuningBeff/g\simeq0.8$. 
    \textbf{\Figb}, In interleaved experimental sequences, we determine expectation values of $\langle \sigma_y \rangle(t)$ as a function of coupling duration $t$.
    We show 242 data points (light blue disks), 600 experimental repetitions per data point, and the corresponding TI simulation (dark blue dotted line).
    We attribute the periodic loss and gain in contrast to time-dependent build-ups of correlations between $S$ and $E$, maximised for $t\simeq \{\T/2,3T/2\}$.
    \textbf{\Figc}, Following this procedure, we determine the normalised Larmor frequency, $\delta\tilde{\omega}(t)/\omega$, as a function of coupling duration from six runs and show averages as light blue disks. The black solid line represents the analytical prediction of energy renormalisation in the JC regime. In contrast, the dark blue dotted line results from TI simulations.
    We observe a maximal shift of $\simeq15 \%$ of the bare spin frequency.
    All error bars represent the standard error of the mean.}
    \label{fig4}
\end{figure}

We employ Ramsey interferometry to measure the frequency of $S$ and observe its shift due to the described renormalisation effect, eventually revealing its time dependence.
We first assess $\omegaTimeAverage$ as a function of $\detuning$ and then resolve $\omegaTimeDependent$ in a close to resonance case. Both sequences are depicted and explained in Fig.\,\hyperref[fig3]{3a} and \hyperref[fig4]{4a}, respectively.

In the first sequence, to probe $\omegaTimeAverage$, we couple $S$ and $E$ for duration $\Tdetuning$.
The sinusoidal model fits the raw data, as illustrated in Fig.\,\hyperref[fig3]{3b}, and yields the accumulated phase of coherences in $S$ and, thus, the average energy shift in $\Tdetuning$.
Figure~\hyperref[fig3]{3c} displays the probed $\omegaTimeAverage$ for a small detuning range, including the JC model prediction equation (\ref{Eq_average_shift}) (black solid line), the TI simulation (dark blue dotted line) and the Lamb shift (grey dotted line). In the dispersive limit $|\detuning|\gg g$, the Lamb shift agrees with $\langle \delta\tilde\omega(t, |\detuning|\gg g) \rangle_{T}$ and is given by $-g^2/\detuning$~\cite{Brune1994,Fragner2008,Colla2024thesis}.
We find agreement between experimental and numerical data (only performed for $\abs{\detuning}/g>0.25$\,\cite{Supplement}), as well as convergence to Lamb shift prediction.
In addition, for $\detuning>0$, the minimal-dissipation ansatz equation (\ref{Eq_average_shift}) agrees on a similar level, while for $\detuning<0$, we find only qualitative agreement.
We attribute these systematic deviations to residual thermal contributions and the impact of a carrier term in the TI Hamiltonian~\cite{Supplement}.
Nonetheless, our results demonstrate the measurement of dressed-state energies near resonance by a systematic phase accumulation encoded in $\S$ -- traditionally used exclusively in the dispersive regime.

In our second sequence, to resolve a time-dependent variation, we choose $\detuningBeff/g\simeq0.8$ and vary the $S$-$E$ coupling duration $t$.
As shown in Fig.\,\hyperref[fig4]{4b}, we provide raw data of six accumulated measurement runs (light blue data points) of the expectation value of $\sigmay(t)$.
We further depict numerical TI simulations of the experimental sequence (dark blue dotted line), which agrees with the raw data.
We observe varying amplitudes indicating correlations with the environment, such as alternating decoherence and recoherence, which result from strong memory effects due to the finite size of $E$.

We evaluate the instantaneous Larmor frequency by counting the zero-crossings of $\sigmay(t)$; this simplified procedure allows us to automatically average out fast oscillating contributions that are due to the counter-rotating terms in the case of the TI Hamiltonian and to keep only the leading renormalisation effect due to the JC interaction, cf. Methods. 
The measured energy splitting, depicted in Fig.\,\hyperref[fig4]{4c}, shows a modulation with maximum variations of about 15$\%$ and an average shift of $\simeq4\%$ of~$\omega$.
For our parameter choice, the variation of the spin energy is related to the build-up of $S$—$E$ correlations. 
Our findings indicate that effects like the Lamb and AC Stark shifts emerge from such correlations, as well as their temporal evolution, which is typically observed only on a time average.
The data points agree with the theoretical prediction made using the method of minimal dissipation~\cite{Colla2022} applied to the JC model, cp. equation \eqref{Eq_shift}, and provide direct evidence of time-dependent renormalisation of the spin energy levels. 
Furthermore, we observe a slight asymmetry in the height of the two recorded dips due to the finite mode temperature with an average motional expectation value $\nth = 0.08(2)$. 
More dramatic changes could, in principle, be observed at higher temperatures; in such a parameter regime, however, the JC Hamiltonian, cp. equation \eqref{JC_ham}, is no longer suitable for the prediction of the TI evolution~\cite{Leibfried2003, clos_time-resolved_2016}.

The time-dependent energy shift recorded here can be considered a generalisation of the well-known time-independent Lamb shift. It can be regarded as resulting from an energy exchange between $\S$ and $\E$, which manifests itself as time-dependent driving terms in the effective Hamiltonian, even though the total Hamiltonian is time-independent.
Our study demonstrates straightforward ways to experimentally probe a system's effective energy levels while it is strongly coupled to its environment. 
It also suggests that the emergent Hamiltonian $K_S$ (following the ansatz of minimal dissipation), which generally exists for any open system, is promising in predicting energy renormalisation in other physical systems.
Since $K_S$ only depends on the open system dynamics, it can, in principle, be retrieved experimentally via process tomography~\cite{Poyatos1997}, generally accessible by any quantum technology platform. 
Further explorations on this line include a better understanding of the emergent Hamiltonian for the analogue spin degree of freedom of a trapped ion when the mode is at higher temperatures and the investigation of the effects of non-thermal initial environmental states (e.g., coherent or squeezed states, or Fock states). 
The emergent driving appearing in these situations could then be exploited for the understanding of the thermodynamical properties of quantum systems, and it allows for testing the prospects of development of autonomous quantum heat engines~\cite{Linden2010}, with the possibility of achieving enhanced efficiencies in non-Markovian scenarios~\cite{Picatoste2023}.
Outside the current platform, the following steps include studying the renormalisation of multiple energy levels and the possible observation of driving due to a non-Markovian but continuous environment.

\bibliography{biblio.bib}

\subsection{Methods}

\subsubsection{Experimental Setup}

Our experimental platform represents an overall nearly perfectly isolated quantum system on all relevant time scales~\cite{clos_time-resolved_2016, Wittemer2018, Hasse_2024_Strobo}, from which we select two appropriate subsystems $S$ and $E$: We choose a single $^{25}\text{Mg}^+$ ion trapped in a linear Paul trap, with a drive frequency of $\simeq2\pi\,56.3$~MHz.
Due to the nuclear spin of $^{25}\text{Mg}$ of $5/2$, the applied homogeneous magnetic quantisation field $\abs{\bf{B_\text{quant}}} \simeq 0.58$~mT induces a hyperfine splitting. We define as pseudo spin (qubit) degree of freedom (DOF) \mbox{$\ket{F=3, m_F = 3} = \ket{\downarrow}$} and \mbox{$\ket{F=2, m_F = 2} =\ket{\uparrow}$}, where $F$ is the total angular momentum and $m_F$ is the projection of the angular momentum along the magnetic field axis, resulting in a transition frequency \mbox{$\omegaasterisk \simeq 2\pi\,1.8$~GHz}.
The phonon DOF is described by three decoupled harmonic oscillators with frequencies \mbox{$2\pi \{1.3, 2.9, 4.5\}$~MHz}.
The lowest mode is oriented approximately along the axial direction, while the two higher modes are oriented in the radial plane.
The cycling transition is tuned near the $S_{1/2}$ and $P_{3/2}$ transition ($\simeq 280$~nm) and is used for Doppler cooling and detection, the regarding $k$-vector of the laser beams are aligned along  $\bf{B_\text{quant}}$.
Additional laser beams are used for repumping and state preparation to couple to appropriate Zeeman substates of $S_{1/2}$ and $P_{1/2}$.
To determine the electronic state population, the fluorescence of the ion is detected via a photomultiplier tube. For further evaluation, we analyse the photon histograms~\cite{Leibfried2003}.
For determining the thermal contribution $\nth$, the population distributions of the motional states are reconstructed by mapping them onto the electronic states~\cite{Leibfried2003}.
For this, we utilise a two-photon stimulated-Raman (TPSR) transitions, which are detuned from the $S_{1/2}$ to $P_{3/2}$ transition by $\simeq + 2\pi\,20$~GHz, and the effective $k$-vector of the used TPSR beam combination coincide with the axial eigenmode orientation enabling carrier and sideband transitions with tunable Rabi rates $\Omega_\text{R}$ between \mbox{$2\pi\,\{0.05, 0.6\}$~MHz} and fixed Lamb-Dicke parameter $\LDp \simeq 0.4$.
Tuning the TPSR beams into the red-sideband (RSB) regime yields a Hamiltonian, that is, in a first-order approximation, formally equivalent to the JC Hamiltonian. 
The axial eigenmode ($E$ degree of freedom) and laser-dressed electronic states ($\S$ degree of freedom) are coupled via the TPSR beams, cp. table~\hyperref[tab:comparisonvalues]{1} and Supplement.
\begin{table*}[ht]
\centering
\begin{tabular}{>{\raggedright}p{0.3\textwidth} >{\raggedright}p{0.3\textwidth} >{\raggedright\arraybackslash}p{0.3\textwidth}}
\hline \hline
    & \textbf{Jaynes-Cummings Model} & \textbf{Trapped-Ion Model (Red-Sideband Regime)} \\ \hline
\textbf{Composition of $S$ and $E$} with characteristic (bare) frequencies $\omega$ and $\omega_m$ & Two-level atom and a single mode of the quantized electromagnetic field (photons) & Laser-dressed states of ion's internal levels and quantized vibrational modes (phonons) \\ \hline
\textbf{Coupling strength} & Direct $ g $ & Laser mediated $ \eta \Omega_\text{R} /2 $ \\ \hline
\textbf{Detuning} $\Delta = \omega_m-\omega$& Depends on the static effective magnetic field $\Beff$, leading to a detuning $\Delta(\Beff)$ & Set by the frequency difference to the first red sideband, $\Delta_{\text{R}}$
 \\ \hline \hline
\end{tabular}
\caption{\textbf{Comparison between the Jaynes-Cummings and trapped-ion (near the red-sideband regime) models.}
More details on this in Ref.~\cite{Leibfried2003, Supplement}.}
\label{tab:comparisonvalues}
\end{table*}
In our setup, we can continuously tune the coupling strength \mbox{$\g/\omega\simeq\{10^{-2}, 10^{-1}\}$} and we choose for the presented experimental results $\g/\omega\simeq0.05$, still below, but close to, the USC regime, cf. Fig.\,\hyperref[fig2]{2c}.

\subsubsection{Determination of Time-Average Shift \label{Sec_Experimental_Sequence_TimeAverageShift}}

\begin{figure}
    \centering
    \includegraphics[]{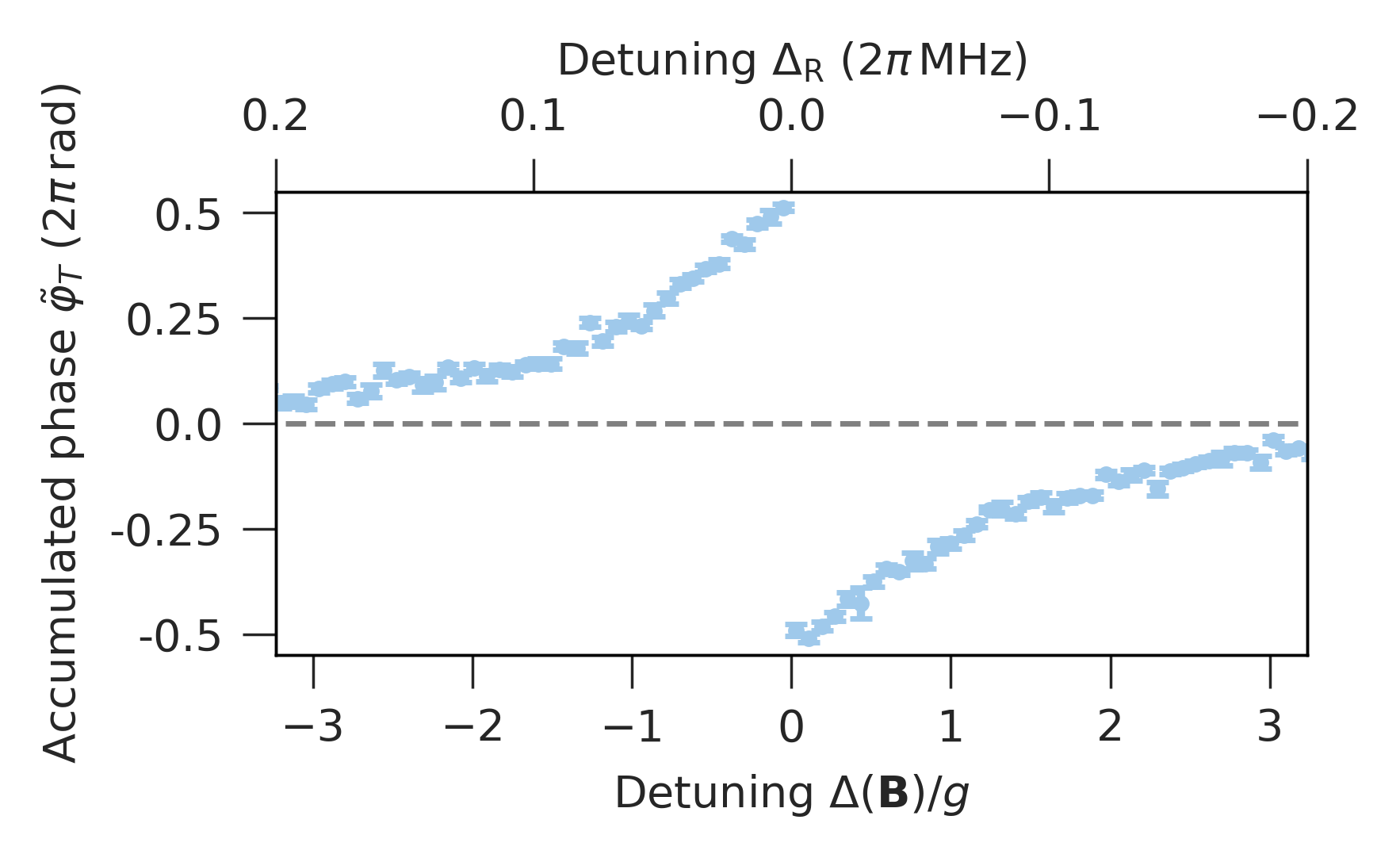}
    \caption{\textbf{Accumulated phase $\accumulatedPhase$ for various detunings $\detuningBeff/g$.}
    We show the sinusoidal fit results of the raw data taken to probe $\omegaTimeAverage$, a necessary analysis step to arrive at Fig.\,\hyperref[fig3]{3c}. Please refer to Section \emph{Determination of Time-Average Shift} for more details.
    Error bars represent the standard deviation of the fit parameter.}
    \label{fig-TimeAverageShift}
\end{figure}

The initial step to determine $\omegaTimeAverage$ is fitting a negative cosine function \mbox{$-C/2 \cos{(\analysisphase + \accumulatedPhase)} + 0.5$}
with the two free parameters contrast $C$ and accumulated phase $\accumulatedPhase$ to the histogram analysed raw data, as exemplarily depicted in Fig.\,\hyperref[fig3]{3b} for three selected $\detuningBeff/g$, with the boundary condition \mbox{$\accumulatedPhase \xrightarrow{\abs{\detuning}\gg0} 0$}.
This boundary condition arises from the fact that \mbox{$\Tdetuning\xrightarrow{\abs{\detuning}\gg0} 0$}, hence the two arms of the Ramsey interferometer are identical, resulting in no expectable phase accumulation.
Note that additional waiting durations are implemented between the $\pi/2$ and $\pi$ pulses, which ensures that independent of the coupling pulse duration $\Tdetuning$, both arms have an equal duration for all $\detuning$. 
The measured accumulated phase, illustrated in Fig.\,\hyperref[fig-TimeAverageShift]{5} is related to $\omegaTimeAverage$ via the equation \mbox{$\accumulatedPhase=\int_{0}^{T} \tilde{\omega}(t) \,dt = \omegaTimeAverageNoShift \cdot T$} with \mbox{$\tilde{\omega}(t)=\omega + \omegaTimeDependent$}.
Following this relations and assuming $\omegaTimeAverageNoShift$ to be static, to obtain $\omegaTimeAverage$, we divide $\accumulatedPhase$ by $\Tdetuning$ and receive the result shown in Fig.\,\hyperref[fig3]{3c}.

\subsubsection{Determination of Time-Dependent Shift/Larmor Frequency}
In our analysis, we use a customised algorithm to detect the frequency of zero crossings in the time series signal of $\langle \sigma_z \rangle (t)$, as shown in Fig.\,\hyperref[fig4]{4b}. Utilising our knowledge of $\omega$, the algorithm clusters adjacent crossings to reduce the impact of fast oscillations or noise. We identify the median times, $T_{\text{zero,}i}$ (where $i$ is the number of zero crossings), within these clusters to determine the zero-crossing timings, where $\langle \sigma_z \rangle (T_{\text{zero,}i}) \simeq 0$. Finally, the time-dependent Larmor frequency values, $\omega_L(t)$, (Fig. \hyperref[fig4]{4c}) are estimated from the timing differences between neighbouring zero crossings using the formula: $\omega_L(t_i) = \pi / (T_{\text{zero,}i+1} - T_{\text{zero,}i})$.

\subsection{Funding}
This project has received funding from the European Union’s Framework Programme for Research and Innovation Horizon 2020 (2014-2020) under the Marie Skłodowska-Curie Grant Agreement No. 847471, the Deutsche Forschungsgemeinschaft (DFG) (Grant No. 217 SCHA 973/6-2), and the Georg H. Endress Foundation.

\subsection{Conflict}
The authors declare no competing interests.

\subsection{Data availability}
All data supporting the plots in this paper and other study findings are available from F.H. on reasonable request.

\subsection{Code availability}
Simulation codes in this paper are available from U.W., and analysis codes from F.H. upon reasonable request.

\subsection{Acknowledgments}
We thank Frederike Doerr for her help checking data evaluation and making comparative measurements.
A.C. would like to thank Edoardo Carnio and Janine Franz for their precious insights.

\subsection{Author contribution}
A.C. developed the theory under the supervision of H.P.B.
F.H. performed the experiment and data analysis under the supervision of T.S. and U.W.
D.P. assisted in preparing the experimental setup and executing the experiment.
U.W. performed numerical trapped-ion and Jaynes-Cummings simulations.
A.C. and F.H. contributed equally to this work and wrote the paper with all authors' input.
\end{document}



\title{Supplemental Material for\\Observing Time-Dependent Energy Level Renormalisation\\ in an Ultrastrongly Coupled Open System}
\author{Alessandra Colla}

\affiliation{Institute of Physics, University of Freiburg, 
Hermann-Herder-Stra{\ss}e 3, D-79104 Freiburg, Germany}

\affiliation{Dipartimento di Fisica ``Aldo Pontremoli'', Universit\`a degli Studi di Milano, Via Celoria 16, I-20133 Milan, Italy}

\author{Florian Hasse}

\affiliation{Institute of Physics, University of Freiburg, 
Hermann-Herder-Stra{\ss}e 3, D-79104 Freiburg, Germany}

\author{Deviprasath Palani}

\affiliation{Institute of Physics, University of Freiburg, 
Hermann-Herder-Stra{\ss}e 3, D-79104 Freiburg, Germany}

\author{Tobias Schaetz}

\affiliation{Institute of Physics, University of Freiburg, 
Hermann-Herder-Stra{\ss}e 3, D-79104 Freiburg, Germany}

\affiliation{EUCOR Centre for Quantum Science and Quantum Computing,
University of Freiburg, Hermann-Herder-Stra{\ss}e 3, D-79104 Freiburg, Germany}

\author{Heinz-Peter Breuer}

\affiliation{Institute of Physics, University of Freiburg, 
Hermann-Herder-Stra{\ss}e 3, D-79104 Freiburg, Germany}

\affiliation{EUCOR Centre for Quantum Science and Quantum Computing,
University of Freiburg, Hermann-Herder-Stra{\ss}e 3, D-79104 Freiburg, Germany}

\author{Ulrich Warring}

\affiliation{Institute of Physics, University of Freiburg, 
Hermann-Herder-Stra{\ss}e 3, D-79104 Freiburg, Germany}

\maketitle


\section{Comparison of Model Descriptions\label{Comparison_Model_Descriptions}}
This section compares two seminal models in quantum mechanics: the Quantum-Rabi (QR) model, its Jaynes-Cummings (JC) approximation, and the Trapped-Ion (TI) model. 
The QR model, integral to quantum optics, describes interactions between a two-level atom and a single electromagnetic field mode. In its simplest form with rotating wave approximation (RWA) it is typically called the JC model. 
The TI model, instrumental in experimental quantum information, simulation, and metrology applications, focuses on well-isolated atomic ions trapped in electromagnetic traps.
Despite their distinct origins, both models, when the TI model is viewed under Lamb-Dicke (LD) and RWA, converge for certain parameter regimes in their descriptions of light-matter interactions and experimentally accessible analogues of it: 
Generally, the trapped-ion platform can be tuned to match the JC dynamics and study analogue features experimentally while using analytical JC model predictions. 
See table~\hyperref[tab:comparison]{S1} for an overview of relevant parameters and approximations.
\begin{table*}
\caption{\label{tab:comparison} 
\textbf{Comparison of the Quantum-Rabi model -- approximated by the Jaynes-Cummings (JC) model -- and the Trapped-Ion model -- approximated in the so-called red-sideband (RSB) and Lamb-Dicke (LD) regime.}
The parameters are defined as follows: $\omega$ (part of the system $S$) and $\omega^*$ (System $S^*$) are the two-level and effectively bare, laser-dressed two-level (pseudo-spin) frequencies respectively, $\omega_m$ (part of system/environment $E$) represents the harmonic oscillator (photon/phonon mode) frequency, $g$ is the coupling strength of the QR model, $\RabiRate/(2\,\pi)$ is the Rabi frequency of the spin-mode coupling in the TI model, $a$ and $a^\dagger$ are the annihilation and creation operators of the modes, $\sigma_-$ and $\sigma_+$ are the lowering and raising operators, $\sigma_z$ are the Pauli $z$-operators, and $\eta$ is the LD parameter. 
The non-linear TI coupling operator $C(\eta, a, a^\dagger) = \exp[\mathrm{i} \eta (a^\dagger + a)]$ simplifies to $C_{\text{LD}}(\eta, a, a^\dagger) = 1+ \mathrm{i} \eta (a^\dagger + a)$ under the Lamb-Dicke approximation (LDA) for $\eta \ll 1$. 
For details on derivations and explanations, see Refs.\cite{wineland_experimental_1997, leibfried_quantum_2003, clos_time-resolved_2016}. In the case of LDA and RWA, both approximated interaction descriptions converge, and we identify $\omega = ((\omega^*)^2+\Omega_R^2)^{1/2}$, and $g = \eta \RabiRate/2$.}
\begin{ruledtabular}
\begin{tabular}{ccc}
 & Quantum-Rabi Model & Trapped-Ion Model\\
\hline
System S/$S^*$ (atom/spin)& $\hbar \omega \sigma_z/2$ & $\hbar \omega^* \sigma_z/2$\\
Evironment $E$ (cavity/phonon mode) & $\hbar \omega_m a^\dagger a$ & $\hbar \omega_m a^\dagger a$\\
Full S/$S^*$-E interactions & $\hbar g (a^\dagger + a) (\sigma_- + \sigma_+)$ & $\hbar\RabiRate/2\,[C(\eta, a, a^\dagger)^\dagger \sigma_- + C(\eta, a, a^\dagger) \sigma_+ ]$\\
\hline
Approx. interactions (JC / RSB)\,\footnote{In the case of the TI model, the approximation is called the red-sideband (RSB) regime\,\cite{wineland_experimental_1997, leibfried_quantum_2003}} & $\hbar g (a^\dagger \sigma_- + a \sigma_+)$ & $ \hbar\eta\RabiRate/2 (a^\dagger \sigma_- + a \sigma_+)$ \\ 
\end{tabular}
\end{ruledtabular}
\end{table*}
We can study and quantify the similarity between the two models using the trace distance of the two-level subsystem using a numerical approach implemented in QuTiP.

\begin{figure}
\includegraphics{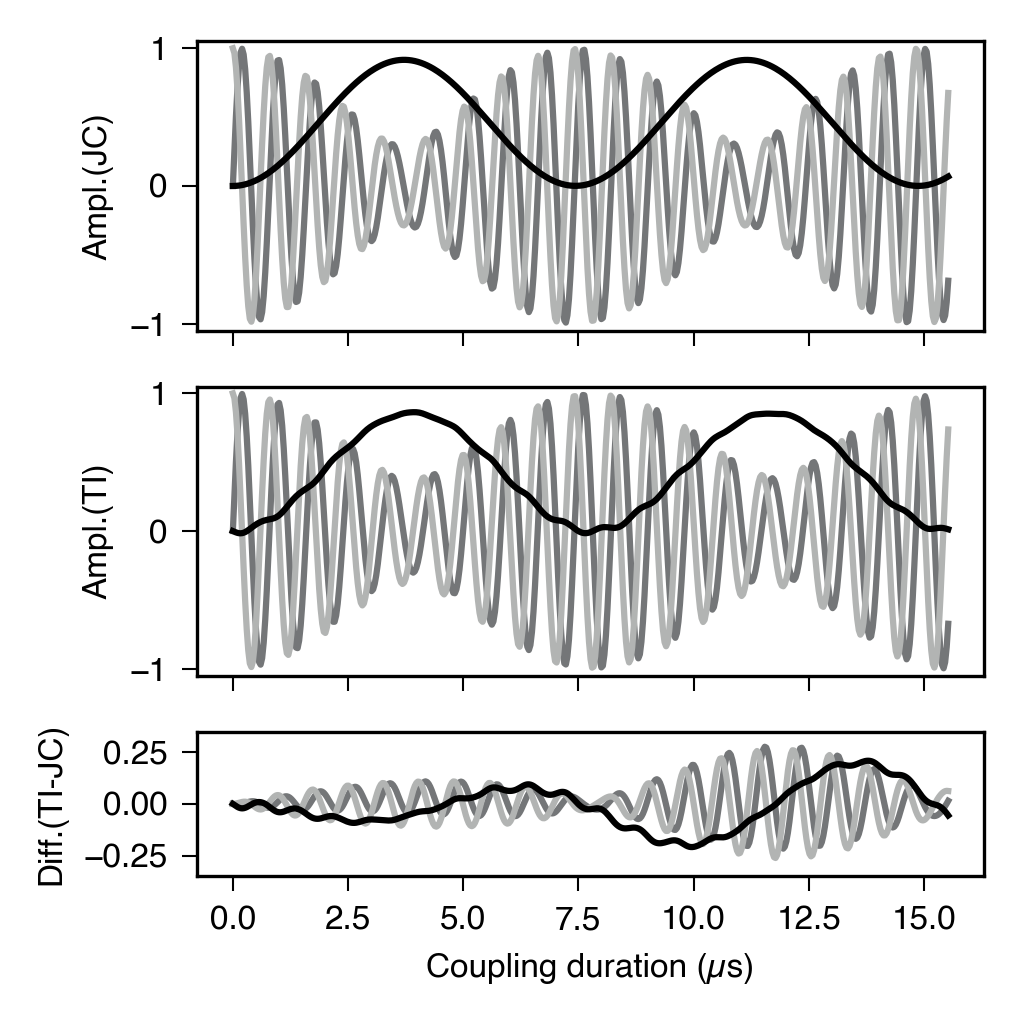}
\caption{\label{figs1}
\textbf{Numerical comparison between the JC and TI models.}
The numerical evolution of expectation values following the JC model (top) and TI model (middle) are shown for different spin components as a function of coupling duration: $\langle \sigma_x \rangle$ in light grey, $\langle \sigma_y \rangle$ in grey, and $\langle \sigma_z \rangle$ in black. Parameters are matched in the numerical simulation except for the modified $\omega$ in JC with respect to the TI $\omega^*$. Differences (bottom graph) in dynamics underscore the effects of fast-rotating and non-linear terms in the TI model.}
\end{figure}

For a comprehensive comparison, we examine the time evolution of $\langle \sigma_x \rangle$, $\langle \sigma_y \rangle$, and $\langle \sigma_z \rangle$ in both models (see Fig.\,\hyperref[figs1]{S1}). The best match between the models is given by taking the coupling parameter in the JC model as \mbox{$g= \eta \Omega_R/2$} and the two-level system level spacing as \mbox{$\omega = ((\omega^*)^2+\Omega_R^2)^{1/2}$}, cf. see table~\hyperref[tab:comparison]{S1}. This is necessary to account for the carrier term, which is present in the TI case, resulting from the first order expansion of the TI coupling in the LD approximation. This term also induces a basis rotation on the spin system, which can be typically neglected for $\omega\gg \Omega_R$. The differences in dynamics found in Fig.\,\hyperref[figs1]{S1} highlight the impact of fast-rotating and non-linear coupling terms in the TI model.

\begin{figure}
\includegraphics{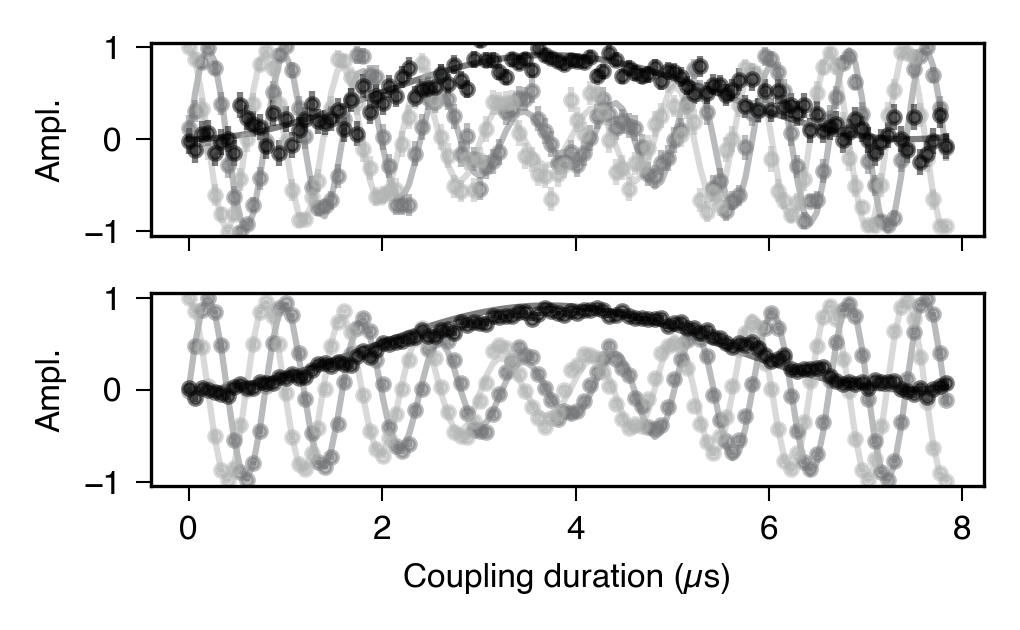}
\caption{\label{figs2}
\textbf{Numerical comparison of the Jaynes-Cummings model ($H^{\text{RWA}}_{\text{JC}}$) and the full Trapped Ion model ($H_{\text{TI}}$) with fundamental quantum projection noise (QPN).} 
We simulate the effect of QPN with 50 and 500 repetitions per data point and compare in total 100 data points with estimates from $H^{\text{RWA}}_{\text{JC}}$ (solid lines).
Expectation values $\langle \sigma_x \rangle$, $\langle \sigma_y \rangle$, and $\langle \sigma_z \rangle$ are displayed in black, grey, and light grey, respectively. Collecting these data sets approximates 5 and 30 minutes of measurement time.
}
\end{figure}

This analysis illustrates the nuanced differences in dynamics between the models, with the TI model's rapid rotations and nonlinearities contrasting the simpler JC dynamics. Repeated measurements are necessary to discern these subtle effects amidst quantum projection noise (QPN)\,\cite{itano_quantum_1993, Wittemer2018} and technical noise contributions. We illustrate the effects of the fundamental QPN for our type of measurement in Fig.\,\hyperref[figs2]{S2} and indicate the required efforts and stability of the system $S$.

%
%

\section{Emergent Hamiltonian for the JC model \label{Hamiltonian_JC}}

The exact master equation for the two-level system in the JC model can be derived following \cite{Smirne2010} for initial environmental states that commute with the number operator. The effective Hamiltonian and thus the time-dependent shift can then be evaluated exactly depending on the initial state of the mode.
For example, for a thermal initial environmental state \mbox{$\rho_E(0) = e^{-\beta \omega_m a^\dag a} /\text{Tr}\{e^{-\beta \omega_m a^\dag a}\}$}, we obtain the following time-dependent frequency shift
\begin{equation}\label{JC-shift-T}
\delta \tilde{\omega}(t) = - \text{Im} \left\{ \frac{\dot{\gamma} (t)}{\gamma (t)} \right\} \; ,
\end{equation}
where
\begin{align}
\gamma(t) = \left(1-e^{-\beta \omega_m}\right)\sum_{n=0}^{\infty} c(n,t)c(n+1,t) e^{-\beta \omega_m n}  \; .
\end{align}
Here we have introduced the coefficients
\begin{align}
  c \left( {n}, t \right)  =&  e^{-i \Delta t / 2}  \left[ \cos \left(
  \Omega_n \frac{t}{2} \right) + i \Delta \frac{\sin \left( \Omega_n
  \frac{t}{2} \right) }{\Omega_n}  \right] \; , 
\end{align}
with $\Omega_n =  \sqrt{\Delta^2 + 4 g^2 {n}}$. Note that the shift equation \eqref{JC-shift-T} depends nontrivially on the initial inverse temperature $\beta$, on the detuning $\Delta$, and on the coupling strength $g$.

If the mode is initially in the vacuum, $\rho_E(0)=\ket{0}\!\!\bra{0}$, the renormalised frequency shift with respect to the bare frequency has the analytical expression
\begin{equation}\label{eq:EX-ren-shift-vacuum}
\delta\tilde{\omega}(t) = - \frac{2g^2}{\Delta} \frac{1}{1+ \frac{\Omega_1^2}{\Delta^2} \cot^2\left(\frac{\Omega_1}{2}t\right)} \; ,
\end{equation}
where we have defined the Rabi frequency \mbox{$\Omega_1 := \sqrt{\Delta^2+4g^2}$}.
Thus, the effective Hamiltonian shows a time-dependent driving of the spin frequency even when the mode is in the ground state. It is periodic (with period \mbox{$T(\Delta)=2\pi/\Omega_1$)} and with a sign dependent on the sign of the detuning $\Delta$. 

The average renormalisation in the case of the vacuum is given by the average of the time-dependent shift over one period. We obtain:
\begin{align}\nonumber
\omegaTimeAverage &= - \frac{\Omega_1}{2\pi} \frac{2g^2}{\Delta} \int_0^{2\pi/\Omega_1} d t \frac{1}{1+ \frac{\Omega_1^2}{\Delta^2} \cot^2\left(\frac{\Omega_1}{2}t\right)}  \\ \label{eq:EX-ren-shift-vacuum-avg}
&=    -\frac{2g^2 \text{sign}(\Delta)}{\Omega_1 + |\Delta|}\; .
\end{align}
In the dispersive limit of large detuning --- namely, \mbox{$|\Delta|\gg g$} --- the average shift becomes
\begin{equation} \label{eq:EX-ren-shift-vacuum-avg-disp}
\omegaTimeAverage \xrightarrow{|\Delta|\gg g} - \frac{g^2}{\Delta} \; ,
\end{equation} 
which is the known Lamb shift for the JC model \cite{Brune1994}

Furthermore, one can check that the time-averaged renormalised frequency in the vacuum case is given exactly by one of the two dressed state energies of the JC Hamiltonian in the one excitation manifold, either the lower or the higher energy depending on the detuning. Indeed it holds
\begin{align}
\omega + \omegaTimeAverage = E_- \theta(\Delta) + E_+ [1-\theta(\Delta)] \; ,
\end{align}
with 
\begin{align}\label{eq:EX-eigenenergies}
E_\pm = (\omega_m +\omega \: \pm \:\Omega_1)/2 \; ,
\end{align}
the dressed energies associated with the eigenstates of the JC Hamiltonian.

\bibliography{biblio}